\documentclass{article}

\usepackage{amssymb} 
\usepackage[colorlinks,linkcolor=green,anchorcolor=red,citecolor=blue]{hyperref}

\begin{document}

\title{The Asymptotic Mandelbrot Law of Some Evolution Networks}

\author{Li~Li}

\maketitle

\begin{abstract}
In this letter, we study some evolution networks
that grow with linear preferential attachment. Based upon some recent results
on the quotient Gamma function, we give a rigorous proof of the
asymptotic Mandelbrot law for the degree distribution $p_k \propto (k + c)^{-\gamma}$
 in certain conditions. We also analytically derive the best fitting values for the scaling
 exponent $\gamma$ and the shifting coefficient $c$.
\end{abstract}

Complex networks are now the joint focus of many branches of research$^{[1-3]}$.
Particularly, the scale-free property of some networks attracts continuous interests, due
 to their importance and pervasiveness$^{[4-6]}$. In short, this property means that
 the degree distribution of a network obeys a power law $P(k) \propto k^{-\gamma}$, where
 $k$ is the degree and $P(k)$ is the corresponding probability density, and the scaling
 exponent $\gamma$ is a constant. A pioneering model that generates power-law degree distribution was presented
by Barab\'{a}si and Albert (BA)$^{[4]}$.

In recent studies, it was found that in some complex networks, e.g. transportation networks$^{[7]}$
and social collaboration networks$^{[8]}$, the degree distribution follows the so-called
 ``shifted power law''$^{[9]}$ $P(k) \propto (k+c)^{-\gamma}$, where the shifting coefficient $c$ is another constant.
 This property is also called ``Mandelbrot law''$^{[10]}$.

To understand the origins of such Mandelbrot law, Ren, Yang and Wang$^{[11]}$ proposed a
interesting growing network that is generated with linear preferential attachment. In
such networks, there exits a recursive dependence relationship between every two
consecutive degrees
\begin{equation}
\label{equ:1}
p(k) \left[ k + \frac{2 + 2 m \beta}{1-\beta}\right] =
p(k-1) \left[ k + \frac{2 m \beta}{1-\beta} - 1 \right]
\end{equation}
where \noindent where $k=2$, ..., $n$, $n$ is the number of nodes.
$m$ is a positive integer constant and $\beta \in [0,1]$ is another constant.

Defining $a = \frac{2 m \beta}{1-\beta} - 1$, $b= \frac{2 + 2 m \beta}{1-\beta}$,
we can abbreviate Eq.(1) as
\begin{equation}
\label{equ:2} p_k \left[ k + b \right] = p_{k-1} \left[ k + a \right]
\end{equation}

To derive the asymptotic of the degree distribution, Ren, Yang and Wang$^{[11]}$
studied the following three kinds of approximations:

I) forward-difference approximation, assuming
\begin{equation}
\label{equ:3} \frac{d p(k)}{d k} \approx p(k) - p(k-1) = p(k) - \frac{k+b}{k+a} p(k) = \frac{a-b}{k+a} p(k)
\end{equation}
\noindent we have an estimation of the power-law as
\begin{equation}
\label{equ:4} p(k) \propto \left( k+a \right)^{-(b-a)}
\end{equation}

II) backward-difference approximation, assuming
\begin{equation}
\label{equ:5} \frac{d p(k)}{d k} \approx p(k+1) - p(k) = \frac{k+1+a}{k+1+b} p(k) - p(k) = \frac{a-b}{k+1+b} p(k)
\end{equation}
\noindent we have another estimation of the power-law as
\begin{equation}
\label{equ:6} p(k) \propto \left( k+b+1 \right)^{-(b-a)}
\end{equation}

III) Suppose we must have a Mandelbrot law $p(k) \propto (k+c)^{-\gamma}$. As a result,
we have $p(k-1) \propto (k-1+c)^{-\gamma}$. Substitute these two approximations
 in the logarithm type of Eq.(2), we have
\begin{equation}
\label{equ:7} \ln \frac{k+a}{k+b} = \ln \frac{p(k)}{p(k-1)} = -\gamma \ln (k+c) + \gamma \ln (k-1+c)
\end{equation}

Rewrite Eq.(7) as
\begin{equation}
\label{equ:8} \ln \frac{1+ a\frac{1}{k}}{1+ b\frac{1}{k}} = \gamma \ln \frac{1+(c-1)\frac{1}{k}}{1+c\frac{1}{k}}
\end{equation}
\noindent and apply the second order Taylor expansion of $\frac{1}{k}$ in Eq.(8), we have
\begin{equation}
\label{equ:9} p(k) \propto \left( k+\frac{b+a+1}{2} \right)^{-(b-a)}
\end{equation}

All these three estimations indicates that the scaling exponent of the degree
distribution should be $-(b-a)$. Simulation results$^{[11]}$ show that Eq.(9) gives
the best approximation accuracy of the empirical distributions. However,
we still need a rigorous proof of this interesting finding.

Indeed, further assuming $\sum_{k=1}^n p(k) = 1$, we have the following matrix
equation
\begin{equation}
\label{equ:10} \left[
\begin{array}{ccccc}
 2+a & -(2+b) & 0 & ... & 0 \\
 0 & 3+a & -(3+b) & ... & 0 \\
  &  & ... & & \\
 0 & 0 & ... & n+a & -(n+b) \\
 1 & 1 & ... & 1 & 1
\end{array} \right] \left[
\begin{array}{c}
 p(1) \\
 p(2) \\
 ... \\
 p(n-1) \\
 p(n)
\end{array} \right] = \left[
\begin{array}{c}
 0 \\
 0 \\
 ... \\
 0 \\
 1
\end{array} \right]
\end{equation}

Using Gaussian elimination algorithm, we can directly solve $p(n)$ from Eq.(10) as
\begin{eqnarray}
\label{equ:11} p(n) & = & \left[1 + \frac{n+b}{n+a} + ... + \prod_{j=2}^n \frac{j+b}{j+a} \right]^{-1} \nonumber \\
& = & \left[ 1+ \sum_{i=2}^n \prod_{j=i}^n \frac{j+b}{j+a} \right]^{-1}
\end{eqnarray}

Based on the recursive relationship Eq.(2), for a given $n$, we have
\begin{eqnarray}
\label{equ:12} p(k) & = & p(n) \left( \prod_{j=k+1}^n \frac{j+b}{j+a} \right) = p(n) \left( \frac{\prod_{j=1}^n \frac{j+b}{j+a}}{\prod_{j=1}^k \frac{j+b}{j+a}} \right) \nonumber \\
& = & p(n) \left( \prod_{j=1}^n \frac{j+a}{j+b} \right) \left( \prod_{j=1}^k \frac{j+a}{j+b} \right)
\end{eqnarray}
\noindent where $k=1$, ..., $n-1$.

It is well known that for Gamma function $\Gamma(z)$, we have $\Gamma(z+1) = z \Gamma(z)$. So, we get
\begin{equation}
\label{equ:13} (j+b) = \frac{\Gamma(j+1+b)}{\Gamma(j+b)}, ~~~(j+a) = \frac{\Gamma(j+1+a)}{\Gamma(j+a)}
\end{equation}
\noindent where $j=1$, ..., $n-1$.

From Eq.(12), we have
\begin{eqnarray}
\label{equ:14} p(k) & = & p(n) \left( \prod_{j=1}^n \frac{j+a}{j+b} \right) \left( \prod_{j=1}^k \frac{\Gamma(j+1+a)}{\Gamma(j+a)} \right) \left( \prod_{j=1}^k \frac{\Gamma(j+b)}{\Gamma(j+1+b)} \right) \nonumber \\
& = & p(n) \left( \prod_{j=1}^n \frac{j+a}{j+b} \right) \frac{\Gamma(k+1+a)}{\Gamma(1+a)} \frac{\Gamma(1+b)}{\Gamma(k+1+b)} \nonumber \\
& = & \lambda \cdot \frac{\Gamma(k+1+a)}{\Gamma(k+1+b)}
\end{eqnarray}
\noindent where $\lambda = p(n) \left( \prod_{j=1}^n \frac{j+a}{j+b} \right) \frac{\Gamma(1+b)}{\Gamma(1+a)}$ is a constant.

Eq.(14) indicates that $p(k)$ has the same asymptotic behavior of $\frac{\Gamma(k+1+a)}{\Gamma(k+1+b)}$.
Actually, the quotient of two Gamma functions is a difficult problem that
received consistent attentions$^{[12-15]}$. There are numbers of approximation
formulas which are not accurate enough for the above applications. Fortunately,
an important results had been obtained very recently$^{[15]}$ as

{\bf Lemma 1$^{[15]}$} Given two constants $s$ and $t$, when $x \rightarrow \infty$, we have
\begin{equation}
\label{equ:15} \left[ \frac{\Gamma(x+t)}{\Gamma(x+s)} \right]^{\frac{1}{t-s}}
\sim \sum_{k=0}^\infty F_k (t,s) x^{-n+1}
\end{equation}
where $F_k(t,s)$ are the polynomials of order $n$ defined by
\begin{equation}
\label{equ:16} F_0 (t,s) = 1
\end{equation}
\begin{equation}
\label{equ:17} F_n(t,s) = \frac{1}{n} \sum_{k=1}^n (-1)^{k+1} \frac{B_{k+1}(t)-B_{k+1}(s)}{(k+1)(t-s)} F_{n-k}(t,s)
\end{equation}
\noindent where $n \ge 1$, $B_k(t)$ is the Bernoulli polynomials (page 40 of ${[16]}$) for $t$.

Based on {\bf Lemma 1}, from Eq.(14), we can have an accurate expansion of the degree distribution as follows
\begin{equation}
\label{equ:18} \left[ \frac{p(k)}{\lambda} \right]^{\frac{1}{a-b}} \sim k + \frac{a+b+1}{2} + \frac{1-(a-b)^2}{24} k^{-1} + ...
\end{equation}

As $k \rightarrow \infty$, we have $\left[ \frac{p(k)}{\lambda} \right]^{\frac{1}{a-b}} \approx k + \frac{a+b+1}{2}$.
Thus, we reach the following conclusion rigorously.

{\bf Theorem 1} The degree distribution follows an asymptotic Mandelbrot law Eq.(9) for some
complex networks that grow with linear preferential attachment depicted by Eq.(2).

\section*{\bf Acknowledgement}
\hskip 7pt
{\footnotesize
We would like to thank Prof. Tao Zhou at School of Computer Science \& Engineering, University of Electronic Science and Technology of China, for drawing our attentions to this problem.
}



\begin{thebibliography}{1}

\bibitem{1}
R. Albert, A. L. Barab\'{a}si, ``Statistical mechanics of complex networks,'' Review of Modern Physiscs, vol. 74, no. 1, pp. 47-97 2002.

\bibitem{2}
S. N. Dorogovtsev, J. F. F. Mendes, ``Evolution of networks: From Biological Nets to the Internet and WWW,'' Advances in Physics, vol. 51, no. 4, pp.  1079-1187, 2002.

\bibitem{3}
M. E. J. Newman, ``The structure and function of complex networks,'' SIAM Review, vol. 45, no. 2, pp. 167-256, 2003.

\bibitem{4}
A.-L. Barab\'{a}si, R. Albert, ``Emergence of scaling in random networks,'' Science, vol. 286, no. 5439, pp. 509-512, 1999.

\bibitem{5}
H.-X. Yang, B.-H. Wang, J.-G. Liu, X.-P. Han, T. Zhou, ``Step-by-Step random walk network with power-law clique-degree distribution,'' Chinese Physics Letters, vol. 25, no. 7, pp. 2718-2720, 2008.

\bibitem{6}
J.-L. Guo, ``Scale-free Networks with Self-Similarity Degree Exponents,'' Chinese Physics Letters, vol. 27, no. 3, id. 038901, 2010.

\bibitem{7}
H. Chang, B.-B. Su, Y.-P. Zhou, D.-R. He, ``Assortativity and act degree distribution of some collaboration networks,'' Physica A, vol. 383, pp.
687-702, 2007.

\bibitem{8}
Y.-L. Wang, T. Zhou, J.-J. Shi, ``Empirical analysis of dependence between stations in Chinese railway network,'' Physica A, vol. 388, no. 14, pp. 2949-2955, 2009.

\bibitem{9}
D.-R. He, Z.-H. Liu, B.-H. Wang, {\it Complex Systems and Complex Networks} (Beijing: Higher Education Press), 2009.

\bibitem{10}
B. Mandelbrot, {\it Information Theory and Psycholinguistics} (New York: Basic Books Publishing Co.), 1965.

\bibitem{11}
X.-Z. Ren, Z.-M. Yang, B.-H. Wang, ``Mandelbrot law of evolution networks,'' Journal of University of Electronic Science and Technology of China, vol. 40. no. 2, pp. 163-167, 2011.

\bibitem{12}
J. S. Frame, ``An approximation to the quotient of Gamma function,'' The American Mathematical Monthly, vol. 56, no. 8, pp. 529-535, 1949.

\bibitem{13}
A. Erd\'{e}lyi, F. G. Tricomi, ``The asymptotic expansion of a ratio of gamma functions,'' Pacific Journal of Mathematics, vol. 1, no. 1, pp. 133-142,
1951.

\bibitem{14}
J. Abad, J. Sesma, ``Two new asymptotic expansions of the ratio of two gamma functions,'' Journal of Computational and Applied Mathematics, vol. 173,
no. 2, pp. 359-363, 2005.

\bibitem{15}
T. Buri\'{c}, N. Elezovi\'{c}, ``Bernoulli polynomials and asymptotic expansions of the quotient of gamma functions,'' Journal of Computational and Applied Mathematics, vol. 235, no. 11, pp. 3315-3331, 2011.

\bibitem{16}
A. Jeffrey, H.-H. Dai, eds., {\it Hanndbok of Mathematical Formulas and Integrals}, 4th edition, Elsevier, Burlington, MA, 2008.

\end{thebibliography}
\end{document}